# Unveiling the polarity of the spin-to-charge current conversion in $Bi_2Se_3$

J. B. S. Mendes[1,*], M. Gamino[2,3], R. O. Cunha[1,4], J. E. Abrão[2], S. M. Rezende[2] and A. Azevedo[2]

[1]*Departamento de Física, Universidade Federal de Viçosa, 36570-900 Viçosa, MG, Brazil*
[2]*Departamento de Física, Universidade Federal de Pernambuco, 50670-901 Recife, PE, Brazil*
[3]*Departamento de Física, Universidade Federal do Rio Grande do Norte, 59078-900 Natal, RN, Brazil*
[4]*Centro Interdisciplinar de Ciências da Natureza, Universidade Federal da Integração Latino-Americana, 85867-970 Foz do Iguaçu, PR, Brazil*

We report an investigation of the spin- to charge-current conversion in sputter-deposited films of topological insulator $Bi_2Se_3$ onto single crystalline layers of YIG ($Y_3Fe_5O_{12}$) and polycrystalline films of Permalloy (Py = $Ni_{81}Fe_{19}$). Pure spin current was injected into the $Bi_2Se_3$ layer by means of the spin pumping process in which the spin precession is obtained by exciting the ferromagnetic resonance of the ferromagnetic film. The spin-current to charge-current conversion, occurring at the $Bi_2Se_3$/ferromagnet interface, was attribute to the inverse Rashba-Edelstein effect (IREE). By analyzing the data as a function of the $Bi_2Se_3$ thickness we calculated the IREE length used to characterize the efficiency of the conversion process and found that $1.2 \text{ pm} \leq |\lambda_{IREE}| \leq 2.2 \text{ pm}$. These results support the fact that the surface states of $Bi_2Se_3$ have a dominant role in the spin-charge conversion process, and the mechanism based on the spin diffusion process plays a secondary role. We also discovered that the spin- to charge-current mechanism in $Bi_2Se_3$ has the same polarity as the one in Ta, which is the opposite to the one in Pt. The combination of the magnetic properties of YIG and Py, with strong spin-orbit coupling and dissipationless surface states topologically protected of $Bi_2Se_3$ might lead to spintronic devices with fast and efficient spin-charge conversion.

*Corresponding author: Joaquim B. S. Mendes, joaquim.mendes@ufv.br

The investigation of new materials with strong spin-orbit coupling (SOC) has improved the means for the generation and detection of spin currents in nonmagnetic materials. This study gave birth to the emergent subfield of spintronics, named spin orbitronics [1-3]. Despite being a subject of interest for many years to the investigation of magnetocrystalline anisotropy, the SOC has been pivotal to the revolution that spintronics has undergone in the last decade. In particular, heavy metals, such as Pt and Pd, have been used as efficient materials for mutual conversion between spin and charge currents via direct and inverse spin Hall effects (SHE and ISHE, respectively) [4-7]. In the last decade, there has been significant progress towards developing materials with strong SOC, which can produce current-driven torques strong enough to switch the magnetization of a ferromagnetic (FM) layer in a spin-valve structure. Such improvement in the SHE has been observed in a wide variety of systems that include enhancement of the SOC driven by surface roughness and volume impurities [8-11], at 2D materials [12] and interfacial effects [13-15].

Indeed, many spintronics-phenomena driven by interface-induced spin-orbit interaction have been extensively investigated over the last few years. For instance, the inverse Rashba-Edelstein effect [16,17] (IREE) was considered for converting spin into charge current [13] in many interface systems [18-25]. Moreover, other materials with outstanding spintronics properties, the topological insulators (TIs), stand out for the mutual conversion between charge and spin due to the large SOC in surface states that locks spin to momentum [26-29]. TIs are a new class of quantum materials that present insulating bulk, but metallic dissipationless surface states topologically protected by time reversal symmetry, opening several possibilities for practical applications in many scientific arenas including spintronics, quantum computation, magnetic monopoles, highly correlated electron systems, and more recently in optical tweezers experiments [30, 31–34]. It is known that in TIs the effects of SOC are maximized because the electron's spin orientation is fixed relative to its direction of propagation. Among the 3D TIs, $Bi_2Se_3$ is a unique material with large bandgap of 0.35 eV and its surface spectrum consists of single Dirac cone roughly centered within the gap [30]. In spite of the fact that the first investigations of spintronics properties of TIs were performed in samples grown by the Molecular Beam Epitaxy (MBE) [29,35], the sputtering deposition technique has been successfully used to grow high quality $Bi_2Se_3$ [23, 28, 36].

The spin Hall angle ($\theta_{SH}$), used to quantify the mutual conversion between spin and charge current, has limited use in systems in which the cross-section of the charge-current-carrying layer is reduced. Owing to the transverse nature of the spin transport phenomena, SHE is a bulk effect occurring within a volume limited by the spin-diffusion length ($\lambda_{sd}$) [15]. For instance, when a 3D spin current density $J_S$ [$A/m^2$] is injected through an interface with high SOC, it generates a 2D charge current density $J_C$ [$A/m$] by means of the IREE. In this case, the ratio $J_C/J_S = (2e/\hbar)\lambda_{IREE}$ defines a length ($\lambda_{IREE}$) that is used as a parameter to measure the efficiency of conversion between spin- to charge current [2,13]. Not only the absolute value of $\lambda_{IREE}$, but also its polarity must be of interest to understand the physics behind the interplay between spin and charge currents.

Here we report an investigation of the spin- to charge current conversion in bilayers of $Bi_2Se_3(t)/YIG(6~\mu m)$, (YIG = $Y_3Fe_5O_{12}$, Yttrium Iron Garnet) by means of the ferromagnetic resonance driven spin pumping (FMR-SP) technique. While the $Bi_2Se_3$ films were grown by DC sputtering, the single-crystal YIG films were grown by Liquid Phase Epitaxy (LPE) onto (111) GGG (=$Gd_3Ga_5O_{12}$) substrates. The pure spin-current density ($J_S$), which flows across the $Bi_2Se_3(t)/YIG$ interface due to the YIG magnetization precession, is converted into a transversal charge current density ($J_C$) that is detected by measuring a DC voltage between two edge contacts. The $Bi_2Se_3$ samples were deposited on top of small pieces of YIG/GGG(111) cut from the same wafer, with thickness 6 μm, width of 1.5 mm and length of 3.0 mm. The YIG films have in-plane magnetization and thus the magnetic proximity effect is expected to shift the Dirac cone sideways along the momentum direction and does not open an exchange gap (i.e. in our heterostructures, the Dirac cone of the TI film will be preserved). The $Bi_2Se_3$/YIG interface has the advantage over the $Bi_2Se_3$/ferromagnetic-metal because it ensures cleaner interface and avoids current shunting as well as spurious spin rectification effects. Previously reported spin-to-charge current conversion experiments with sputtered $Bi_2Se_3$/YIG were carried out in YIG grown by sputtering or MBE and, to the authors knowledge, there is no investigation about the polarity of $\lambda_{IREE}$ [23, 28].

X-ray diffraction (XRD) analysis was carried out by means of out-of-plane scan as well as grazing incidence X-ray diffraction (GIXRD), which is more valuable for assessing ultra-thin film structures. Fig. 1(a) shows the out of plane XRD θ-2θ scan pattern of the $Bi_2Se_3$(6 nm)/YIG(6μm)/GGG sample over a 2θ range between 20° and 70°. The pattern shown in Fig. 1(a) displays reflections associated with the (222) and (444) crystal planes of YIG, proving that the present YIG film is epitaxially grown on the GGG substrate. In the inset, we can see the XRD spectrum at high resolution detailing the double peak corresponding to the (444) Bragg reflections of the GGG substrate and the epitaxial YIG in the (444) plane. In order to optimize the scattering contribution from the $Bi_2Se_3$ films, we used grazing incidence X-ray diffraction (GIXRD) for investigating the $Bi_2Se_3$/YIG(6μm)/GGG samples. As shown in Fig. 1(b), the GIXRD data evidenced the diffraction peaks characteristic of the $Bi_2Se_3$ 6 nm thick film, meaning that the film is polycrystalline and has a preferential texture oriented in the planes: (0 0 9), (0 0 15), (0 0 18), (0 0 21), which is in agreement with the literature [37, 38]. Figure 1(c) shows the X-ray reflectivity (XRR) data for $Bi_2Se_3$(16.0 nm)/Si. The well-defined and the good periodicity of the Kiessig fringes allow an accurate determination of the thickness of $Bi_2Se_3$ films. Figure 1(d) shows an atomic force microscopy (AFM) image of the YIG film surface and confirms the uniformity of the YIG film surface with very small roughness (~0.2 nm). On the other hand, Fig. 1(e) shows the AFM image of the sputtered granular bismuth selenide thin film ($t = 4$ nm) grown onto YIG/GGG substrate. The image shows that $Bi_2Se_3$ film grown onto YIG favors the formation of a granular film, with grain sizes up to ~ 0.3 μm, and has a root-mean-square (RMS) surface roughness of about 1.0 nm. The typical energy-dispersive x-ray (EDX) spectrum of $Bi_2Se_3$ on the YIG film can be seen in Fig. 1(f). The EDX spectrum taken from an arbitrary region of the sample shows the presence only of yttrium (Y), iron (Fe), oxygen (O) of the YIG film; bismuth (Bi) and selenium (Se) of $Bi_2Se_3$. The additional peak of the carbon (C) in the EDX spectrum is due to the presence of carbon tape used as support on which the samples are prepared for analysis. In the figure there are also the EDX-maps showing that the Bi and Se are evenly distributed over the entire surface of the film. Different regions of the samples were analyzed, in order to confirm the results of the EDX measurements.

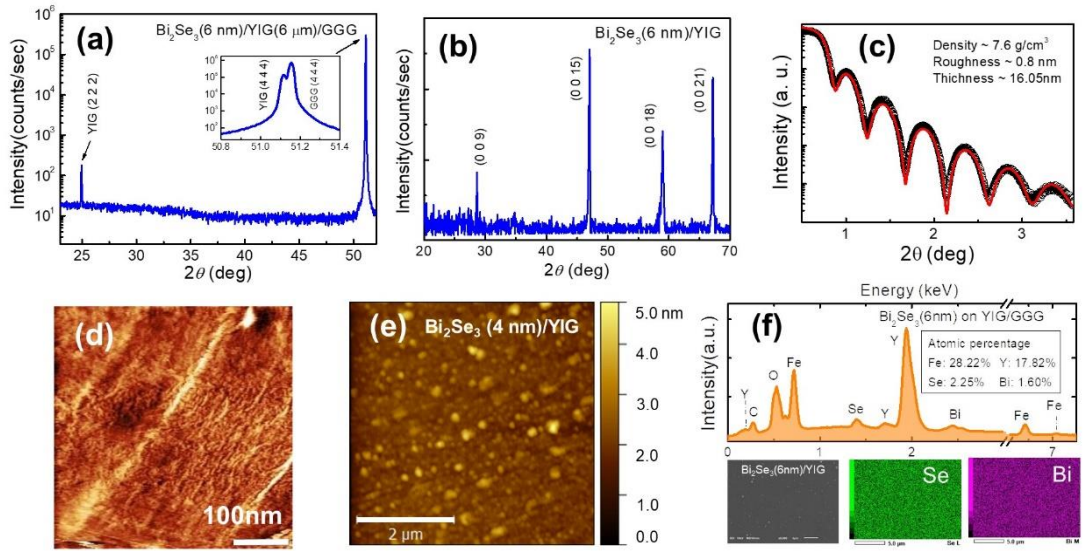

**Fig. 1 (color online).** (a) Out-of-plane XRD patterns (θ-2θ scans) of $Bi_2Se_3$ film grown on YIG/GGG substrate. The XRD spectrum at high resolution detailing the positions of the peaks of the YIG film and the GGG substrate is shown in the inset. (b) The GIXRD pattern of the $Bi_2Se_3$(6 nm)/YIG/GGG sample. (c) XRR spectra of the $Bi_2Se_3$ thin film (t ≈ 16 nm). The red solid line across the XRR data indicates the best fitting obtained for the thickness calibration. (d) AFM image of the YIG film surface. (e) AFM image of the surface of the $Bi_2Se_3$(4nm)/YIG(6μm)/GGG sample. (f) EDX spectrum (top) from an arbitrary region in the sample of $Bi_2Se_3$(6nm)/YIG and EDX-maps (bottom) showing that the Bi and Se are evenly distributed over the entire surface of the film.

Figure 2 (a) illustrates the performed experiments of FMR-SP in which the sample with electrodes at the edges is mounted on the tip of a polyvinyl chloride (PVC) rod and is inserted, via a hole drilled at the bottom wall of a shorted X-band waveguide, in a position of maximum rf magnetic field and zero electric field. The loaded waveguide is placed between the poles of an electromagnet that applies a DC magnetic field $\vec{H}_0$ perpendicular to the in-plane RF magnetic field, $\vec{h}_{rf}$. Electric contacts of silver were sputtered at the edges perpendicular to the larger sample size, so that the spin pumping voltage ($V_{SP}$) can be directly measured by means a nanovoltmeter. As the DC and RF magnetic fields are perpendicular to each other, the sample, attached to a goniometer, can be rotated so that we can investigate de angular dependence of both the ferromagnetic resonance (FMR) as well as $V_{SP}$. Field scan spectra of the derivative $dP/dH$, at a fixed frequency of 9.5 GHz, are obtained by modulating the field $\vec{H}_0$ with a small sinusoidal field at 1.2 kHz and using lock-in amplifier detection. Figure 2 (b) shows the FMR spectrum of a bare YIG sample (3.0 mm x 1.5 mm x 6.0 μm) obtained with the in-plane field applied normal to the larger length with an incident power of 54 mW. The strongest line corresponds to the uniform FMR mode ($k_0 \cong 0$) in which the frequency is given by the Kittel's equation $\omega_0 = \gamma\sqrt{(H_0 + H_A)(H_0 + H_A + 4\pi M_{eff})}$, where $\gamma = 2\pi \times 2.8\ GHz/kOe$ and $4\pi M_{eff} = 4\pi M + H_s \cong 1760\ G$ for YIG. While the lines to the left of the uniform mode correspond to hybridized standing spin-wave surface modes, the lines to the right correspond to the backward volume magnetostatic modes with quantized wave number $k$, subjected to the appropriated boundary conditions. All modes have similar half-width-half-maximum linewidth (HWHM) of $\Delta H_{YIG} = 1.4$ Oe. As shown in Fig. 2(c), the deposition of a 4.0 nm thick film of Bi₂Se₃ on the YIG layer increases the FMR linewidth to $\Delta H_{Bi_2Si_3/YIG} = 1.7$ Oe. This linewidth increase is mostly due to the spin pumping process that transports spin angular moment out of the YIG layer [39,40]. As the YIG magnetization vector precesses, it injects a pure spin current density $\vec{J}_S$, that flows perpendicularly to the YIG/Bi₂Se₃ interface with transverse spin polarization $\hat{\sigma}$, which is given by

$$\vec{J}_S = \left(\hbar g_{eff}^{\uparrow\downarrow}/4\pi M_s^2\right)\left(\vec{M}(t) \times \partial\vec{M}(t)/\partial t\right), \qquad (1)$$

where $M_s$ and $M(t)$ are the saturation and time dependent magnetization, respectively, and $g_{eff}^{\uparrow\downarrow}$ is the real part of the spin interface mixing conductance, that takes into account the forward and backward flows of the spin current [39]. It is important to mention that $J_S$ in Eq. (1) has units of (angular moment)/(time.area). As previously mentioned, $\vec{J}_S$ results in an increased magnetization damping due to the outflow of the spin angular moment, and due to the IREE it generates a transverse charge current in the Bi₂Se₃ film. From the additional linewidth broadening, we can estimate the value of the spin mixing conductance $g_{eff}^{\uparrow\downarrow}$ of the Bi₂Se₃/YIG interface. As $g_{eff}^{\uparrow\downarrow}$ is proportional to the additional linewidth broadening, i.e., $g_{eff}^{\uparrow\downarrow} = (4\pi M t_{FM}/\hbar\omega)(\Delta H_{Bi_2Se_3/YIG} - \Delta H_{YIG})$, where $\omega = 2\pi f$ and $t_{FM}$ is the ferromagnetic (FM) layer thickness for thin FM films (or the coherence length for films such as the used here), and considering that for the Pt/YIG bilayer obtained with the same YIG, $\Delta H_{Pt/YIG} - \Delta H_{YIG} = 0.55$ Oe and $g_{eff}^{\uparrow\downarrow}(Pt/YIG) = 10^{14} \text{cm}^{-2}$, we obtain $g_{eff}^{\uparrow\downarrow}(Bi_2Se_3/YIG) \approx 5.4 \times 10^{13} \text{cm}^{-2}$.

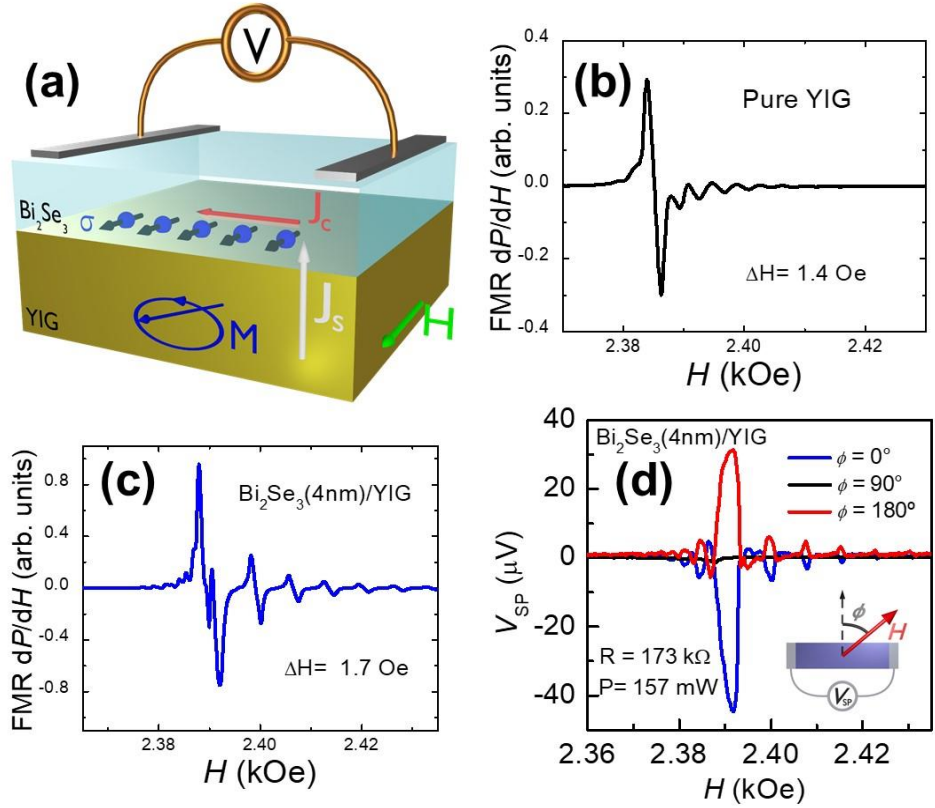

**Figure 2 (color online).** (a) Schematics of the FMR-SP technique in which we highlights the spin-current to charge-current conversion process at the interface. Field scan FMR absorption derivative, for a bare YIG film with thickness of 6 μm (b) and (c) the bilayer of Bi$_2$Se$_3$ (4 nm)/ YIG(6 μm). (d) Field scan of the spin pumping voltage measured for the bilayer Bi$_2$Se$_3$ (4 nm)/ YIG(6 μm) at three different in-plane angles as illustrated in the inset, with an incident microwave power of 157 mW.

The measurement of $V_{SP}$ is carried out by sweeping the DC field with no AC field modulation and directly measuring $V_{SP}$ that is generated between the two electrodes due to the spin-to-charge current conversion. Fig. 2(d) shows the spin pumping voltage, measured directly by a nanovoltmeter, in a bilayer of Bi$_2$Se$_3$(4 nm)/YIG as function of the applied field for three in-plane directions given by $\phi = 0°, 90°$ and $180°$, as illustrated in the inset. As expected from the equation $\vec{J}_C = \theta_{SH}(2e/\hbar)(\vec{J}_S \times \hat{\sigma})$, where $\hat{\sigma} \parallel \vec{H}$, the charge current flows in-plane so that the value of $V_{SP}$ is maximum for $\phi = 0°$ and $\phi = 180°$ for blue and red curves, respectively. While it is null for $\phi = 90°$, as shown by the black curve. The asymmetry between the positive and negative peaks is similar to that observed in other bilayer systems and can be attributed to a thermoelectric effect [41].

While Fig. 3(a) shows the field scans of $V_{SP}$ for $39\ mW \leq P_{rf} \leq 157\ mW$, Fig. 3(b) shows the RF-power dependence of the peak voltage measured at $\phi = 180°$. The linear dependence of the $V_{SP}$ as a function of $P_{rf}$ confirms that we are exciting the FMR in the linear regime. On the other hand, the dependence of the peak voltage as a function of the Bi$_2$Se$_3$ layer thickness ($t_{Bi_2Se_3}$) exhibits a more challenging behavior. It decreases as $t_{Bi_2Se_3}$ increases in a clear opposition with results shown by materials in which the spin- to charge current conversion occurs in the bulk, as in Pt, for example. This decrease in the peak voltage was also observed in crystalline Bi$_2$Se$_3$ grown by MBE [35]. We could try to explain the origin of the voltage in Bi$_2$Se$_3$/YIG as due to the spin pumping ISHE mechanism, by means spin diffusion model where the spin pumping voltage is given by [42-44],

$$V_{SP}(H) = \frac{R_N e \theta_{SH} \lambda_N w p_{xz} \omega g_{eff}^{\uparrow\downarrow}}{8\pi} \tanh\left(\frac{t_N}{2\lambda_N}\right)\left(\frac{h_{rf}}{\Delta H}\right)^2 L(H - H_R) \cos \phi. \tag{2}$$

Here, $R_N$, $t_N$, $\lambda_N$ and w, are respectively the resistance, thickness, spin diffusion length and width of the Bi$_2$Se$_3$ layer, considering the microwave frequency $\omega = 2\pi f$, and $p_{xz}$ is a factor that expresses the ellipticity and the spatial variation of the rf magnetization of the FMR mode. Also, $h_{rf}$ and $\Delta H$ are the applied microwave field and FMR linewidth, and $L(H - H_R)$ represents the Lorentzian function. By assuming that $2\lambda_N \gg t_N$, thus $\tanh(t_N/2\lambda_N) \approx t_N/2\lambda_N$. Therefore, Eq. (2) can be written as $V_{SP} = \left(R_N f e \theta_{SH} w p_{xz} g_{eff}^{\uparrow\downarrow} t_N/8\right)\left(h_{rf}/\Delta H\right)^2$. This expression does not depend on $\lambda_N$, as expected for TIs, so that $t_N$ can be interpreted as an effective thickness attributed to the Bi$_2$Se$_3$. From the measured quantities for the bilayer $Bi_2Se_3(4\ nm)/YIG$, $R_N = 173\ k\Omega$, $g_{eff}^{\uparrow\downarrow} \approx 5.4 \times 10^{13} cm^{-2}$, $h_{rf} = 0.055\ Oe$, $\Delta H = 1.7\ Oe$, $V_{SP} = 44.7\ \mu V$ and $\theta_{SH} \cong 0.11$ [as reported in Ref. [27] for average value of $\theta_{SH}$], the effective thickness of the Bi$_2$Se$_3$ layer is $t_N = 0.46$ Å. This small value is certainly unphysical for an effective layer that converts a 3D spin current density in a 3D charge current, as happens in the SHE effect. However, it provides an evidence that the spin-to-charge current conversion is dominated by surface states of the sputtered Bi$_2$Se$_3$ layer.

To further verify that the spin- to charge-current conversion in $Bi_2Se_3/YIG$ is dominated by the surface states, we can calculate the effective length $\lambda_{IREE} = (\hbar/2e)J_C/J_S$, where $V_{IREE} = R_{Bi_2Se_3} w J_C$ and $J_S = \left(e\omega p_{xz} g_{eff}^{\uparrow\downarrow}/16\pi\right)\left(h_{rf}/\Delta H\right)^2 L(H - H_R)$, with $p_{11} = 0.31$, see Ref. [45]. Therefore, $\lambda_{IREE}$ is given by,

$$\lambda_{IREE} = \frac{4V_{IREE}}{R_{Bi_2Se_3} e w f g_{eff}^{\uparrow\downarrow} p_{xz} \left(h_{rf}/\Delta H\right)^2}. \tag{3}$$

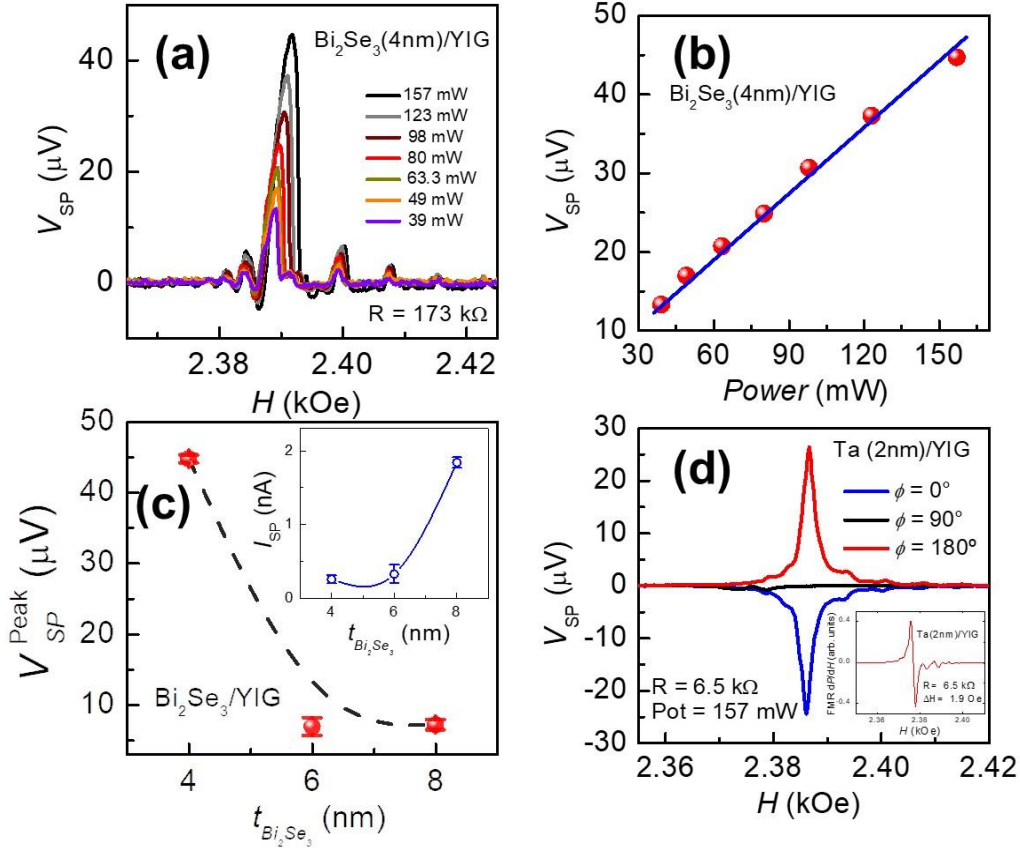

**Figure 3 (color online).** (a) Field scans of $V_{SP}$ for several values of the incident microwave power. (b) Peak voltage value as a function of the incident microwave power measured for the bilayer of Bi$_2$Se$_3$(4 nm)/YIG. (c) Peak voltage value measured as a function of the Bi$_2$Se$_3$ thickness for an incident power of 157 mW. The inset shows the dependence of the spin pumping current ($I_{SP} = V_{SP}^{peak}/R$). (d) Field scans of $V_{SP}$ for the bilayer of Ta(2nm)/YIG obtained at same experimental configuration used to measure $V_{SP}$ in Bi$_2$Se$_3$/YIG. By comparing Fig. 3(d) with Fig. 2(d) we concluded that the $V_{SP}$ polarization of Bi$_2$Se$_3$ is the same as in Ta.

Using the physical quantities for the bilayer $Bi_2Se_3(4\text{ nm})/YIG$, given above, we obtained $|\lambda_{IREE}|(t_{Bi_2Se_3} = 4 \text{ nm}) = (2.2 \pm 0.4) \times 10^{-12}$ m. For the other two bilayers we obtained, $|\lambda_{IREE}|(t_{Bi_2Se_3} = 6 \text{ nm}) = (2.0 \pm 0.5) \times 10^{-12}$ m, and $|\lambda_{IREE}|(t_{Bi_2Se_3} = 8 \text{ nm}) = (1.2 \pm 0.1) \times 10^{-12}$ m. Where only three parameters varied from sample to sample, which are: resistance (R), average voltage $<V_{SP}>$, and the FMR linewidth $\Delta H_{Bi_2Se_3/YIG}$. The error bars were incorporated in $\lambda_{IREE}$ by taking into account the variation of $V_{SP}$ measured at $\phi = 0°$ and 180°. Therefore, we found values that varies in the range of $0.012 \text{ nm} \leq |\lambda_{IREE}| \leq 0.022 \text{ nm}$, and in the literature there are values reported in the range of $0.01 \text{ nm} < \lambda_{IREE} \leq 0.11 \text{ nm}$ [23,35]. Although we cannot rule out the spin diffusion mechanism, the values of $\lambda_{IREE}$ strongly support the role played by the surface states in the spin- to charge-current conversion process

occurring in sputtered $Bi_2Se_3$ layers. Indeed, granular $Bi_2Se_3$ films grown by sputtering keep the topological insulator properties even in the nanometer size regime. The basic mechanisms explaining the existence of topological surface states in granular films of $Bi_2Se_3$ is based on the electron tunneling between grain surfaces. Also, the electron quantum confinement in nanometer sized grains, has been considered as the reason of the high charge-to-spin conversion effect in granular TIs [36].

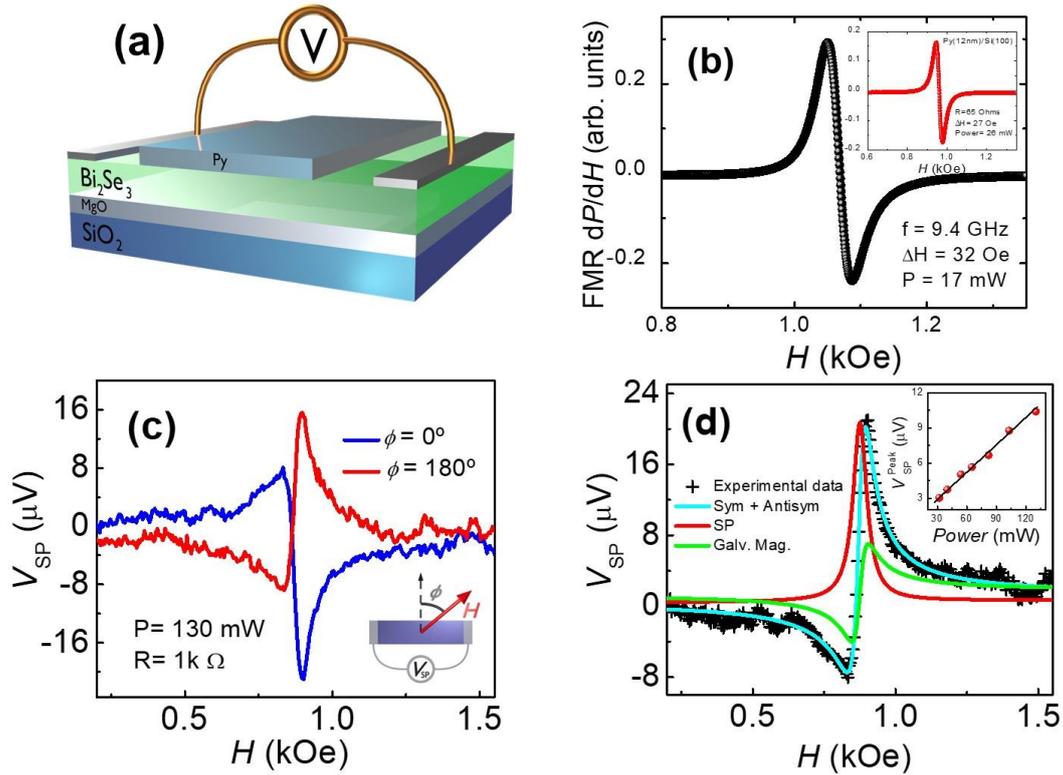

**Figure 4 (color online).** (a) Sketch of the bilayer sample $Py/Bi_2Se_3$. In order to minimize shunting effects, the Py film partially covers the $Bi_2Se_3$ film surface. (b) Derivative FMR field scan for the bilayer of Py (12 nm)/$Bi_2Se_3$ (4 nm) obtained by inserting the sample in a microwave rectangular cavity operating at 9.4 GHz. The inset shows the derivative FMR absorption field scan for the bare Py (12 nm) film. The increase of the linewidth (HMHM) for the bilayer Py (12 nm)/$Bi_2Se_3$ (4 nm) is mostly due to the spin pumping process. (c) Field scan of $V_{SP}$ for two in-plane angles, measured at the same experimental configuration. The result confirms that the sign of the $V_{SP}$ in the bilayer Py (12 nm)/$Bi_2Se_3$ (4 nm), is the same as the one measured in $Bi_2Se_3$/YIG. (d) Decomposition of the symmetric and antisymmetric components of $V_{SP}$ obtained by fitting the data of (c). The inset shows the dependence of the peak value of the symmetric component as a function of the microwave power, measured at $\phi = 180°$.

To further study the polarization of the spin-to-charge current conversion process in $Bi_2Se_3$, we investigated the spin-pumping voltage in the bilayer of $Bi_2Se_3$(4 nm)/Py(12 nm), where Py is Permalloy ($Ni_{81}Fe_{19}$). The investigated sample is illustrated in Fig. 4(a), where the layer of Py partially covers the $Bi_2Se_3$ surface, so that the electrodes are attached out of the Py layer. The sample was sputter grown onto SiO$_2$(300nm)/Si(001) where a passivation layer of MgO(2nm) was grown underneath the Bi$_2$Se$_3$ layer. Fig. 4(b) shows the FMR spectrum of the bilayer Py(12 nm)/Bi$_2$Se$_3$(4 nm) in which the sample is placed in a microwave cavity resonating at 9.4 GHz, with $Q \approx 2000$ and an incident microwave power of 17 mW. Due to the spin pumping effect, the FMR linewidth (HWHM) increased to 32 Oe in comparison with the linewidth of 27 Oe of a bare Py(12 nm) layer, shown in the inset of Fig. 4(b). Figure 4(c) shows the spin pumping voltage measured between the electrodes for $\phi = 0°$ (blue curve) and $\phi = 180°$ (red curve), with an incident power of 170 mW. The $V_{SP}$ lineshape is described by the sum of symmetric and antisymmetric components, $V_{SP}(H) = V_S(H - H_R) + V_{AS}(H - H_R)$, where $V_S(H - H_R)$ is the (symmetric) Lorentzian function and $V_{AS}(H - H_R)$ is the (antisymmetric) Lorentzian derivative centered at the FMR resonance field ($H_R$). Figure 4(d) shows the corresponding symmetric (red) and antisymmetric (green) components of the $V_{SP}$ line shape for $\phi = 180°$, obtained by fitting the data (black symbols) with a sum of a Lorentzian function and Lorentzian derivative (given by the cyan curve). The inset of Fig. 4(d) shows the linear dependence of the peak value of the symmetric component as a function of the incident power. The symmetric component of $V_{SP}$ in Py/Bi$_2$Se$_3$, which is attributed to the spin-to-charge current conversion process, has the same polarity as the one observed for $V_{SP}$ measured in Bi$_2$Se$_3$/YIG and Ta/YIG bilayers.

In conclusion, we report an investigation of the spin- to charge-current conversion process in bilayers of YIG/Bi$_2$Se$_3$ and Py/Bi$_2$Se$_3$, where the Bi$_2$Se$_3$ layer was grown by sputtering. The results obtained by means of the ferromagnetic resonance driven spin pumping technique has shed light in some aspects not investigated by previous papers. We discovered that the spin- to charge-current mechanism in topological insulator Bi$_2$Se$_3$ has the same polarity as the one of Ta, and opposite to the one in Pt. By interpreting the spin pumping voltage as due to the inverse Rashba-Edelstein effect, we calculated the value of $\lambda_{IREE}$ as a function of Bi$_2$Se$_3$ thickness and the values found demonstrate that the surface states have a dominant role in the spin-charge conversion process. Thus, the spin-charge conversion mechanism based on the spin diffusion process plays a secondary role. We expect that our results will be useful for applications in spintronic devices and understanding the spin- to charge-current mechanism in sputter-deposited films of topological insulator Bi$_2$Se$_3$.

This research was supported by Conselho Nacional de Desenvolvimento Científico e Tecnológico (CNPq), Coordenação de Aperfeiçoamento de Pessoal de Nível Superior (CAPES), Financiadora de Estudos e Projetos (FINEP), Fundação de Amparo à Ciência e Tecnologia do Estado de Pernambuco (FACEPE), Fundação Arthur Bernardes (Funarbe), Fundação de Amparo à Pesquisa do Estado de Minas Gerais (FAPEMIG) - Rede de Pesquisa em Materiais 2D and Rede de Nanomagnetismo.